\documentclass[twocolumn,amsmath,amssymb,aps,prb]{revtex4-2}
\usepackage[colorlinks,urlcolor=blue,citecolor=blue,linkcolor=blue]{hyperref}
\usepackage{bm}
\usepackage{graphicx}
\usepackage{subfigure}
\usepackage{epstopdf}
\usepackage{float}
\usepackage{braket}

\begin{document}

\title{The failure of semiclassical approach
in the dissipative fully-connected Ising model}

\author{Ni Zhihao}
\altaffiliation{These authors contributed equally to this work.} 
\affiliation{Department of Physics, Zhejiang Normal University, Jinhua 321004, People's Republic of China}

\author{Qinhan Wu} 
\altaffiliation{These authors contributed equally to this work.} 
\affiliation{Department of Physics, Zhejiang Normal University, Jinhua 321004, People's Republic of China}

\author{Qian Wang}
\affiliation{Department of Physics, Zhejiang Normal University, Jinhua 321004, People's Republic of China}

\author{Gao Xianlong}
\affiliation{Department of Physics, Zhejiang Normal University, Jinhua 321004, People's Republic of China}

\author{Pei Wang}
\email{wangpei@zjnu.cn}
\affiliation{Department of Physics, Zhejiang Normal University, Jinhua 321004, People's Republic of China}

\begin{abstract}
We solve the fully-connected Ising model in the presence of dissipation and
time-periodic field, with the corresponding Lindblad equation having a time-periodic
Liouvillian. The dynamics of the magnetizations is studied by using both the semiclassical
approach and the numerical simulation with the help of permutation symmetry.
The semiclassical approach shows a transition from the periodic response
for small field amplitude to the chaotic dynamics for large amplitude. The trajectory
of the magnetizations and the Lyapunov exponents are calculated, which support
the existence of a chaotic phase. But in the exact
numerical simulation, the response is periodic for both small and large amplitude.
The scaling analysis of Floquet Liouvillian spectrum confirms the periodic response
in the thermodynamic limit. The semiclassical approximation is found to fail
as the field amplitude is large.
\end{abstract}

\maketitle

\section{Introduction}

Dissipative spin models are currently attracting wide interest, because they
describe genuine nonequilibrium states of matter and at the same time,
can be realized in various platforms from superconducting circuits to Rydberg atoms
\cite{puri2017quantum,clerk2020hybrid,xu2020probing,henriet2020quantum,zeiher2017coherent,labuhn2016tunable,saffman2010quantum,angerer2018superradiant,zhang2017observation}.
The model consists of an ensemble of spins that are subject to dissipation caused
by external baths. The dynamics is described by the Lindblad equation,
obtained by integrating out the baths' degrees of freedom~\cite{2002theory}.

Great efforts have been taken in the investigation of various spin models.
In the central spin model~\cite{kessler2012dissipative}, the dissipative phase
transition was located according to the closing of the Liouvillian gap.
The transverse-field Ising model was thoroughly studied both under
the mean-field approximation and beyond~\cite{lee2011antiferromagnetic,lee2012collective,
ates2012dynamical,hu2013spatial,marcuzzi2014universal,weimer2015variational,weimer2015variational}, in which
the bistability of steady states in some region of the parameter space was found
to be replaced by a first-order phase transition after the spatial correlation
was correctly considered. 
As the couplings between spins are all-to-all and then the system can be
seen as a huge spin, the semiclassical approach was adopted.
The magnetization was found to display an everlasting oscillation
in the thermodynamic limit, indicating that the time translational symmetry
is spontaneously broken into a discrete one~\cite{ie2018boundary,shammah2018open,tucker2018shattered}. If the dissipation acts in the eigenbasis of the transverse
field, then a continuous dissipative phase transition manifests itself as
continuous order parameters with discontinuous derivatives~\cite{overbeck2017multicritical,foss2017solvable,jin2018phase,wang2021dissipative}. The XYZ-Heisenberg model was also studied by different approximation schemes~\cite{lee2013unconventional,jin2016cluster,rota2017critical,
casteels2018gutzwiller,huybrechts2020validity} to clarify its phase diagram.

These studies focused on the time-independent Liouvillians. But
much less is known as the Liouvillian changes periodically with time~\cite{zhu2019dicke}.
The topic of this work
is to study the response to a time-periodic Liouvillian.
On the other hand, in closed quantum systems, the response to a time-periodic Hamiltonian
has been under intensive investigations.
The kicked top models were studied both theoretically~\cite{haake1987classical,wang2021multifractality,
bhosale2017signatures,bandyopadhyay2004entanglement,
ruebeck2017entanglement,lombardi2011entanglement,wang2004entanglement,iwaniszewski1995chaos} and
experimentally\cite{chaudhury2009quantum,krithika2019nmr,neill2016ergodic}, which
is known to exhibit a transition between a regular dynamical phase and a chaotic one,
depending on the value of the kicking strength. Similar chaotic behavior was
found in the Lipkin-Meshkov-Glick model as the parameters change
periodically with time~\cite{engelhardt2013ac,lerose2019prethermal,
das2006infinite,russomanno2015thermalization}.
The study in this paper can be seen as an investigation
of the dissipation effect on the chaotic dynamics in the periodically-driven spin models.

As a concise example, we study the fully-connected Ising model in the presence of
collective dissipation. Without dissipation, chaotic behavior
appears in the presence of a strongly oscillating
external field~\cite{lerose2019prethermal,das2006infinite,
russomanno2015thermalization}. We find that, the chaotic behavior
is robust against weak dissipation, if the semiclassical approximation is taken.
As the oscillating amplitude of field increases, a periodic response
changes into a subharmonic oscillation, and then into
a chaotic behavior. But the numerical simulation shows
that, beyond the semiclassical approximation,
only the periodic response can survive the quantum fluctuation, but
neither the subharmonic nor the chaotic dynamics can be observed.
The semiclassical approximation works only as the oscillating amplitude
is small in the periodic-response regime,
but it fails as the amplitude is large.

The paper is organized as follows. We introduce the model
in Sec.~\ref{sec:model}. Section~\ref{sec:semi} contributes to
the discussion of the semiclassical results. The exact numerical
simulations of the dynamics of magnetizations
are discussed in Sec.~\ref{sec:finite}. The Floquet Liouvillian spectrum
is studied in Sec.~\ref{sec:spec}. Finally, Sec.~\ref{sec:con} summarizes our results.

\section{ THE MODEL}
\label{sec:model}

We consider the transverse field Ising model with all-to-all couplings,
and a sinusoidal modulation added to the external field. The Hamiltonian is written as
\begin{equation}\label{1}
\hat{H} =-Ng(\hat{J}_{x})^{2}+N\Gamma(t)\hat{J}_{z},	
\end{equation}
where $N$ denotes the total number of spins, $g$ denotes the coupling strength, and
$\hat{J}_{\alpha}=\sum_{j}\hat{\sigma}_{j}^{\alpha}/N$ denote the collective spin operators
with $\hat{\sigma}_{j}^{\alpha}$ being the Pauli matrices of the $j$th spin and $\alpha=x,y,z$.
$\Gamma(t)$ denotes the time-dependent external field, which is supposed to be
$\Gamma(t)= \Gamma_0+A\ sin(\omega_0 t)$, where $A$, $\Gamma_0$ and $\omega_0$
are the oscillating amplitude, mean value and frequency, respectively.
We set $g = 1$ as the unit of energy throughout the paper.

In the presence of dissipation, the dynamics of the system is
described by the Lindblad equation~\cite{LindbladmasterE}. The density matrix satisfies
\begin{equation}\label{2}
\frac{d\hat{\rho}}{\partial t}=-i[\hat{H},\hat{\rho}]+N{\gamma_{c}}\left(
2\hat{J}_{-}\hat{\rho}\hat{J}_{+}- \{\hat{\rho},\hat{J}_{+}\hat{J}_{-}\}\right),
\end{equation}
where $ \hat{J}_{\pm}=\left(\hat{J}_x \pm i \hat{J}_y\right)/2$ are the jump operators,
and $\gamma_{c}$ is the dissipation rate. As $A=0$ and then $\Gamma(t)= \Gamma_0$
is a constant, the solution of Eq.~\eqref{2}
was studied previously~\cite{wang2021dissipative}. The jump operator forces the spins to be
aligned in the negative $z$-direction, while the interaction between spins favors
an alignment in the $x$-direction. Their interplay results in a steady state, which
is either ferromagnetic (with nonzero magnetization in the $x$-direction)
or paramagnetic. Here we extend to the case of $A\neq 0$, in which
the field oscillation prevents a steady state being reached,
and then we expect a nontrivial dynamical behavior.

\begin{figure}
	
	\subfigure[]{
		\begin{minipage}{0.9\linewidth}
			\centering
			\includegraphics[width=\linewidth]{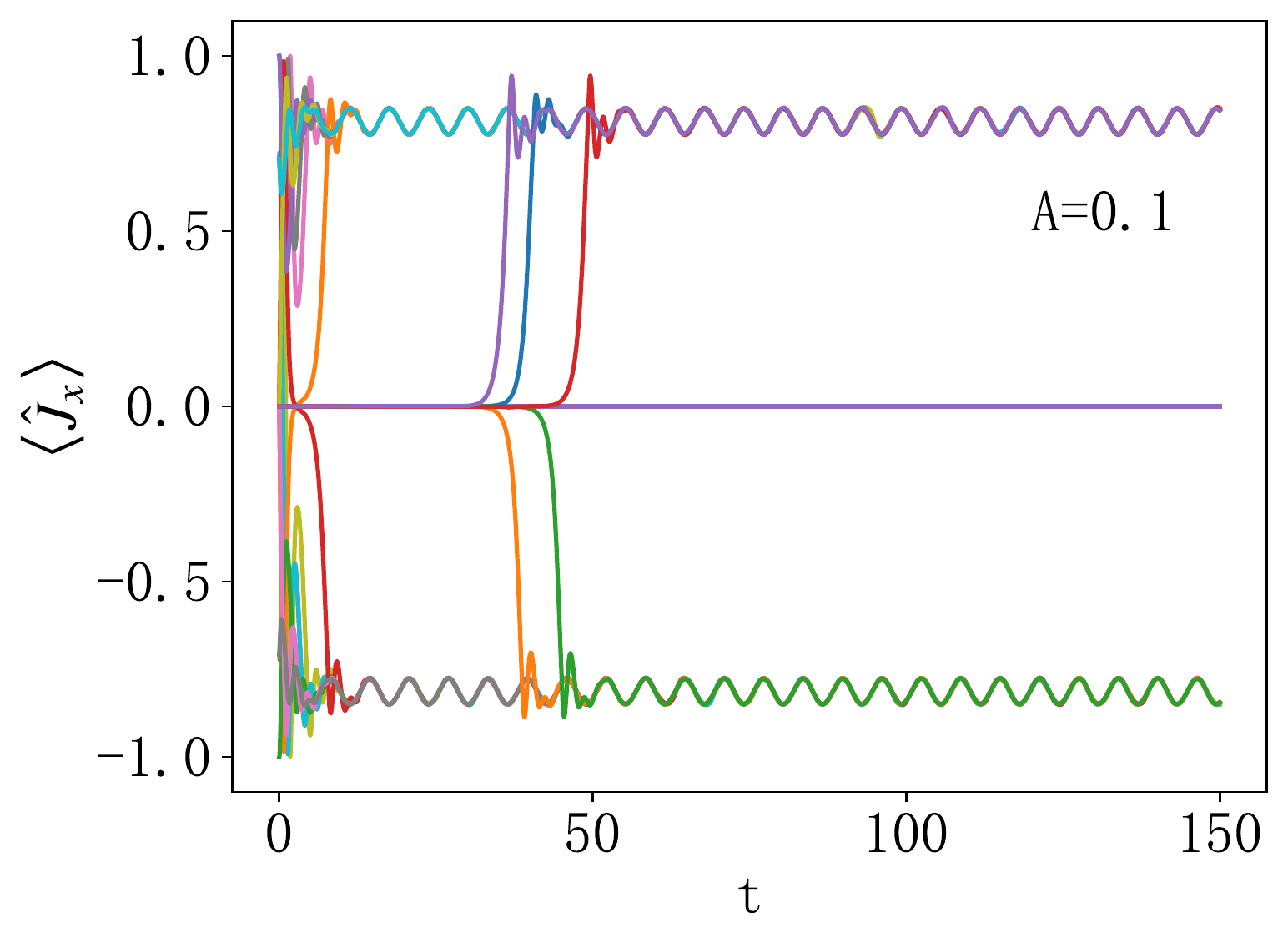}
		\end{minipage}\label{img2a}
	}%

	\subfigure[]{
		\begin{minipage}{0.9\linewidth}
			\centering
			\includegraphics[width=\linewidth]{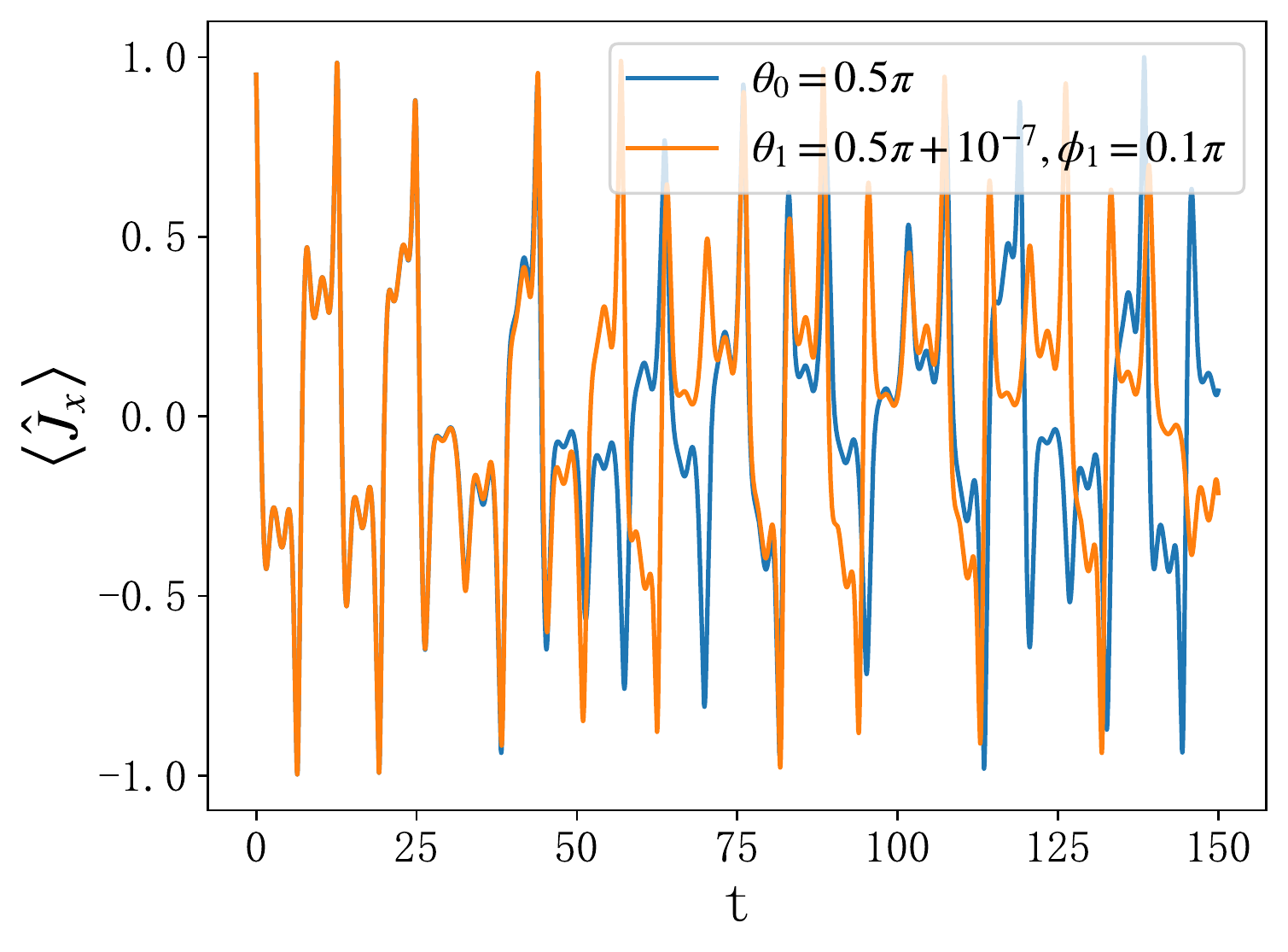}
		\end{minipage}\label{img2b}
	}%
	
	\centering
	\caption{The time evolution of $m_x$ for (a) $A=0.1$ with $25$ different initial states,
	and (b) $A=1$ with two initial states - $(\theta_0=0.5\pi,\phi_0=0.1\pi)$
	and $(\theta_1=0.5\pi+10^{-7},\phi_1=0.1\pi)$. Different line colors
	are for different initial states. The dissipation strength is set to $\gamma_c=1$.}
	\label{img2}
\end{figure}

\begin{figure}
\subfigure[]{
\begin{minipage}{0.9\linewidth}
\centering
\includegraphics[width=\linewidth]{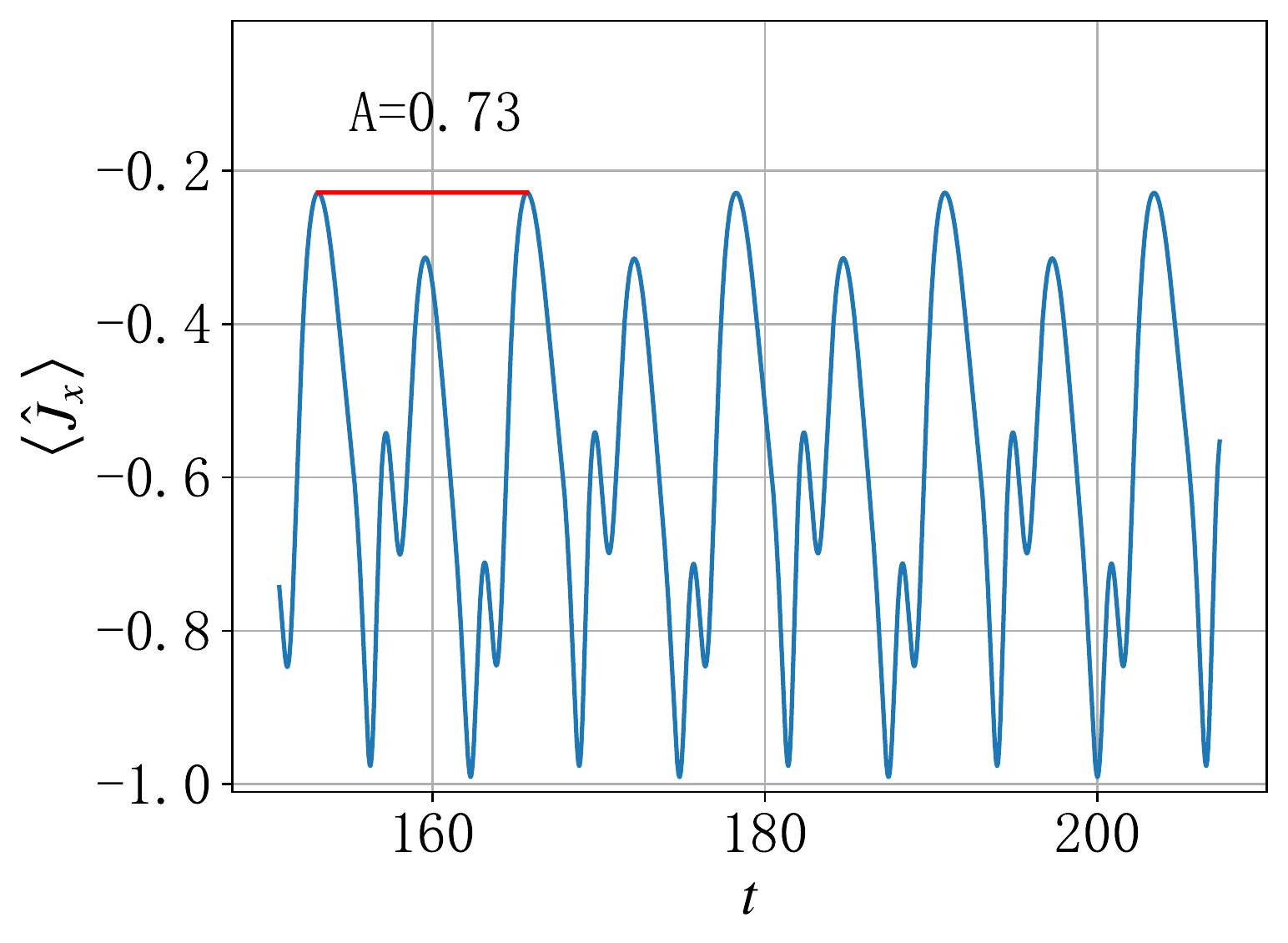}
\end{minipage}\label{img1a}
}
\subfigure[]{
\begin{minipage}{0.9\linewidth}
\centering
\includegraphics[width=\linewidth]{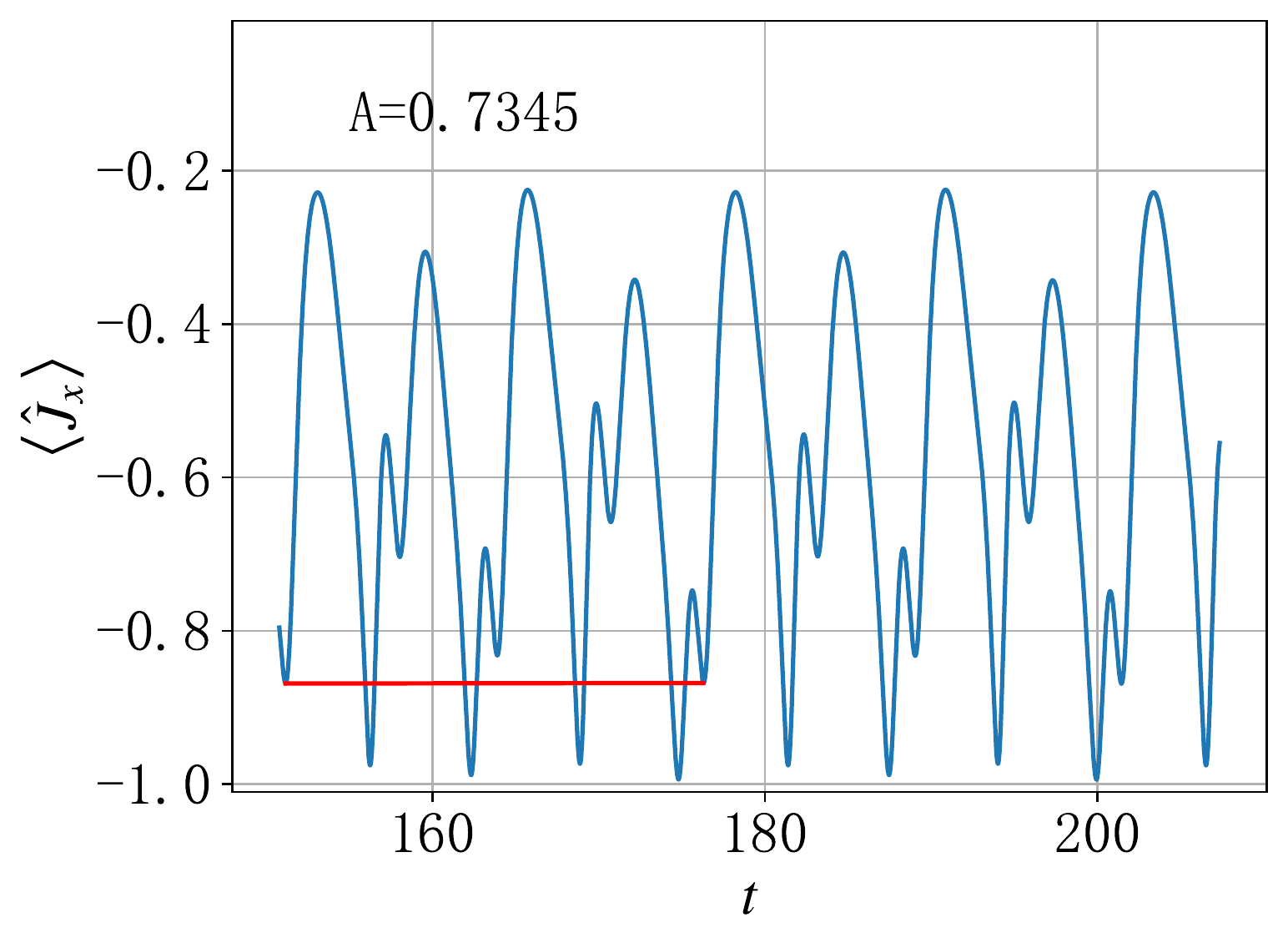}
\end{minipage}\label{img1b}
}

\centering
\caption{The time evolution of $m_x$ for (a) $A=0.73$ and (b) $A=0.7345$.
The red line highlights a complete period. The transient regime
($t<150$) is omitted.}
\label{img1}
\end{figure}

\section{Chaotic dynamics in the semiclassical approach}
\label{sec:semi}

The semiclassical approach is frequently employed for solving the fully-connected models.
We choose the order parameters to be $m_{\alpha}=\left\langle \hat{J}_{\alpha}\right\rangle =Tr[\hat{\rho}\hat{J}_{\alpha}]$. By ignoring the correlations (i.e., setting $\left\langle \hat{J}_{\alpha}\hat{J}_{\beta}\right\rangle =\left\langle \hat{J}_{\alpha}\right\rangle \left\langle \hat{J}_{\beta}\right\rangle$)
in the limit $N\to\infty$, we
obtain a nonlinear system of differential equations, which read
\begin{equation}
\begin{aligned}
\dot{m}_{x}	&=-2\Gamma(t)m_{y}+\gamma_{c}m_{x}m_{z} , \\
\dot{m}_{y}	&=4m_{x}m_{z}+2\Gamma(t)m_{x}+\gamma_{c}m_{y}m_{z}, \\
\dot{m}_{z}	&=-4m_{x}m_{y}-\gamma_{c}(m_{x}^{2}+m_{y}^{2}).
\end{aligned}\label{3}
\end{equation}
It is easy to see that $\left| {\bf{m}}(t)\right|  =\sqrt{m_x^2 + m_y^2 + m_z^2} $
is a constant of motion, which can be set to unity without loss of generality.
${\bf{m}}$ is moving on a Bloch sphere, and the initial state can be
described by the azimuthal angles $\left(\theta, \phi\right)$, which are defined
by $m_x = \sin\theta \cos\phi$, $m_y=\sin\theta\sin \phi$ and $m_z=\cos \theta$.

Since the coefficient $\Gamma(t)$ is a periodic function of $t$ with the period $2\pi/\omega_0$,
one may naively think that the solution ${\bf{m}}(t)$ is also a periodic function. This
is the case for small $A$, but not true for large $A$. We note that Eq.~\eqref{3} bears some
resemblance to the Lorenz equation~\cite{lorenz1963deterministic},
one famous example of the deterministic chaos in the classical systems.
In Eq.~\eqref{3}, the possibility of chaos comes from the fact that $\Gamma(t)$
is time-periodic, even the trajectory is limited on a two-dimensional sphere.
Indeed, we observed a periodic ${\bf{m}}(t)$ for small $A$, but a chaotic
${\bf{m}}(t)$ for large $A$.

Next we choose $\Gamma_0=1$ and $\omega=1$ as an example for the demonstration.
In Fig.~\ref{img2}(a), we display the evolution of
$m_x $ for small $A$ ($A=0.1$) and 25 initial states that are evenly distributed
on the Bloch sphere. Except for the initial state at the south pole (${\bf{m}}=(0,0,1)$),
all the others eventually evolve into one of the two oscillation
modes that are symmetric to each other.
The two oscillation modes have the same period, which is exactly
the driving period ($2\pi/\omega_0$). For small $A$, the long-time response
is insensitive to a small deviation in the initial state.
On the other hand, Fig. \ref{img2b} shows the evolution of $m_x $
for a large $A$ ($A = 1.0$). Till the largest evolution time that is accessible,
no periodicity is observed. More importantly, the long-time response is
extremely sensitive to the initial condition. Even for a tiny deviation
in the initial state ($\theta_1-\theta_0=10^{-7}$), $m_x(t)$
displays a significant difference as $t$ is as large as a few hundreds
(see the lines of different colors in Fig.~\ref{img2b}).

For the values of $A$ between $0.1$ and $1.0$, ${{m}_x}(t)$ displays abundant dynamical
behaviors. As $A$ increases up to a certain critical value, the time period is doubled.
In Fig.~\ref{img1a}, we plot ${{m}_x}(t)$ for $A=0.73$. It is easy to see that
the period is not $2\pi/\omega_0$, instead, it becomes $4\pi/\omega_0$.
As $A$ increases further, the period-doubling happens again.
For example, for $A=0.7345$, the period becomes $8\pi/\omega_0$ (see Fig.~\ref{img1b}).
The period-doubling bifurcation is well-known in the classical nonlinear systems.
Usually, a cascade of period-doubling bifurcations lead to chaos~\cite{galatzer1997understanding}.
This explains why we observe a chaotic dynamics as $A$ is as large as $A=1$.

Depending on the values of $A$, the system exhibits the periodic, subharmonic
or chaotic responses. We plot the trajectory of the vector ${\bf{m}}(t)$ on the Bloch sphere
in Appendix~\ref{sec:app}, which provides more evidence for the existence
of different dynamical behaviors.

In our model, the subharmonic response (i.e., doubled period) to a time-periodic Liouvillian
must be distinguished from that in the Floquet time-crystals.
As will be shown next, the subharmonic response in our model
can only be observed in the semiclassical limit. It cannot
survive the quantum fluctuation that is unavoidable at finite $N$.

\begin{figure}
	\centering
	\includegraphics[width=1\linewidth]{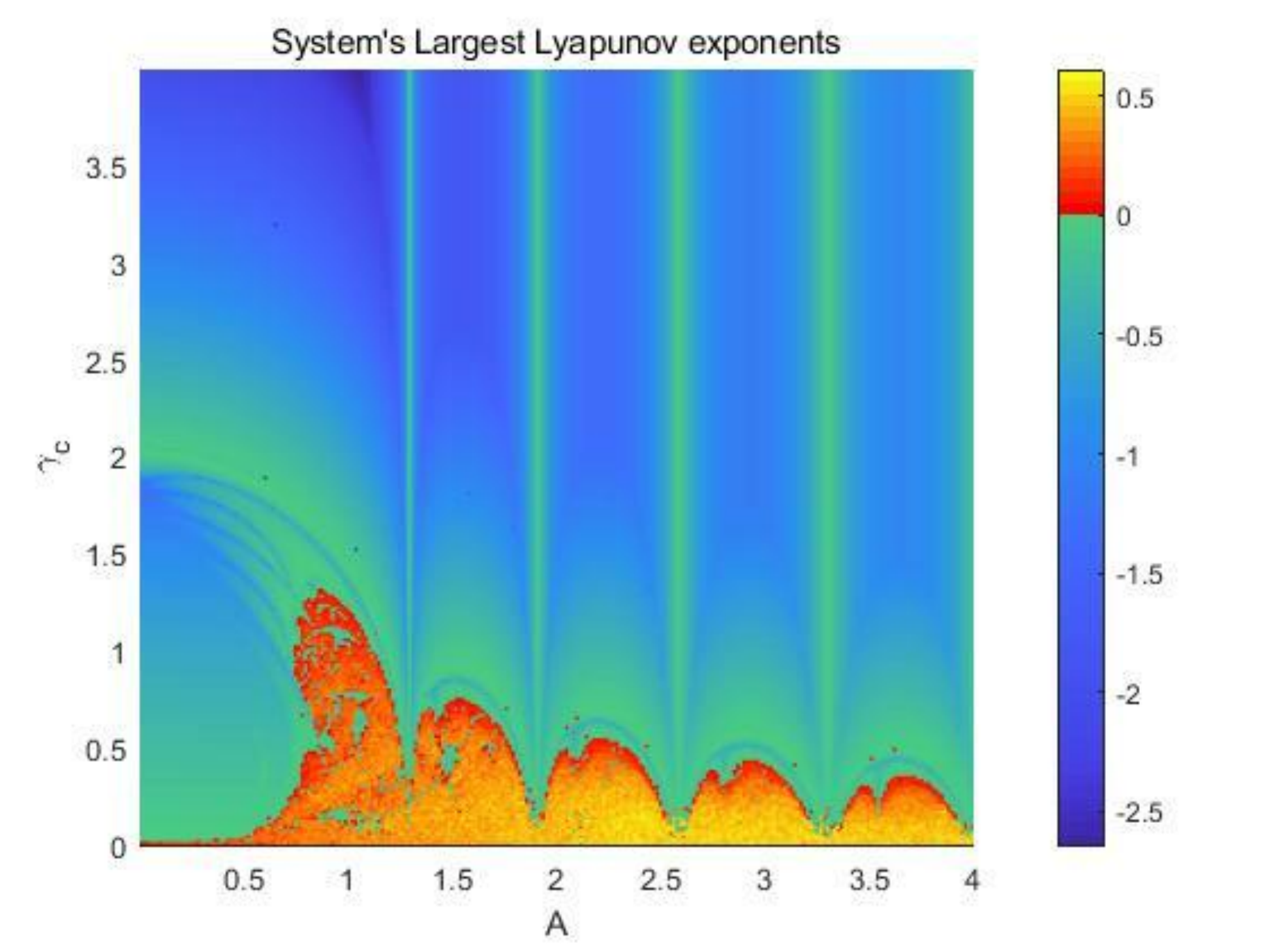}
	\caption[]{The largest Lyapunov exponent as a function of $\gamma_c$ and $A$.
	The system displays chaotic dynamics as $LLE > 0$ (see the area in red or yellow).}
	\label{fig:h0}
\end{figure}

To locate the chaotic phase in the parameter space, we calculate the
Lyapunov exponent. The defining property of a chaotic system is
the extreme sensitivity of trajectories to the initial condition.
Two points that are initially close will drift apart exponentially over time.
The Lyapunov exponent~\cite{oseledets1968multiplicative} provides a quantitative
measure for this, which is defined as
the average exponential rate of convergence or divergence
between adjacent trajectories in the phase space.
Especially, the largest Lyapunov exponent (LLE) is frequently employed
for determining whether a nonlinear system is chaotic.
If the LLE is greater than zero, two initial points will depart
from each other exponentially, and then the system is chaotic~\cite{wolf1985determining},
otherwise, it is not. Figure~\ref{fig:h0} displays the dependence of the LLE on $A$ and $\gamma_c$.
The area in red or yellow has a positive LLE, while the area in blue or green has a negative LLE. 
The chaotic phase is clearly distinguishable in the parameter space.
An interesting finding is that for a fixed $\gamma_c$ in the interval $\left(0,1\right)$,
the dynamics is chaotic only for an intermediate amplitude of oscillating field, but the dynamics
is regular either if $A$ is too large or too small. In the presence of
strong dissipation ($\gamma_c>1.5$), there is no chaotic dynamics for
whatever $A$.

\section{Periodic behavior at finite $N$}
\label{sec:finite}

To check the validity of semiclassical approximation,
we numerically simulate the real-time dynamics at finite $N$, by exploiting
the permutation symmetry of fully-connected models.
The Dicke basis with maximum angular momentum is defined as~\cite{sciolla2011dynamical}
\begin{equation}
\label{4}
\ket{M} =\sqrt{\frac{1}{C_{N}^{\frac{N}{2}+M}}}
\sum_{{\sum_{j=1}^{N} \sigma_{j}=M}}\ket{\sigma_{1},\sigma_{2},\cdots,\sigma_{N}} ,
\end{equation}
where $C_{N}^{\frac{N}{2}+M}$ is the binomial coefficient, and $\sigma_j=\pm 1/2$
represents the spin-up and down states, respectively.
And $M= -N/2, -N/2+1,\cdots, N/2$ is the magnetization in the $z$-direction.
The initial state is supposed to be a pure state with all the spins
aligned in the same direction. We use the azimuthal angles $(\theta,\phi)$
to indicate the initial direction, then the initial state can be
expressed in the Dicke basis as
 \begin{equation}
 \label{key}
 	\ket{ \theta ,\phi } =\sum_{M=-\frac{N}{2} }^{\frac{N}{2}} 
	C_{N}^{M+\frac{N}{2} }\cos\frac{\theta}{2}^{\frac{N}{2}+M}
	\sin\frac{\theta}{2}^{\frac{N}{2}-M}e^{i(\frac{N}{2}-M)\phi}\left | M  \right \rangle .
\end{equation} 
Equation~\eqref{2} has the permutation symmetry, its solution can then be expressed as
$\hat{\rho}(t) = \sum_{M,M'} \rho_{M,M'}(t) \ket{M}\bra{M'}$, where
$\rho_{M,M'}(t)$ is the density matrix in the Dicke basis. Now Eq.~\eqref{2} changes into
a system of ordinary differential equations for $\rho_{M,M'}$, which are solved numerically.
The permutation symmetry reduces the dimension of Hilbert space from $2^N$
to $N$, and then allows us to access a large system with the number of spins $N\sim 10^2-10^3$.

\begin{figure}
	\subfigure[]{
		\begin{minipage}{1\linewidth}
			\centering
			\includegraphics[width=\linewidth]{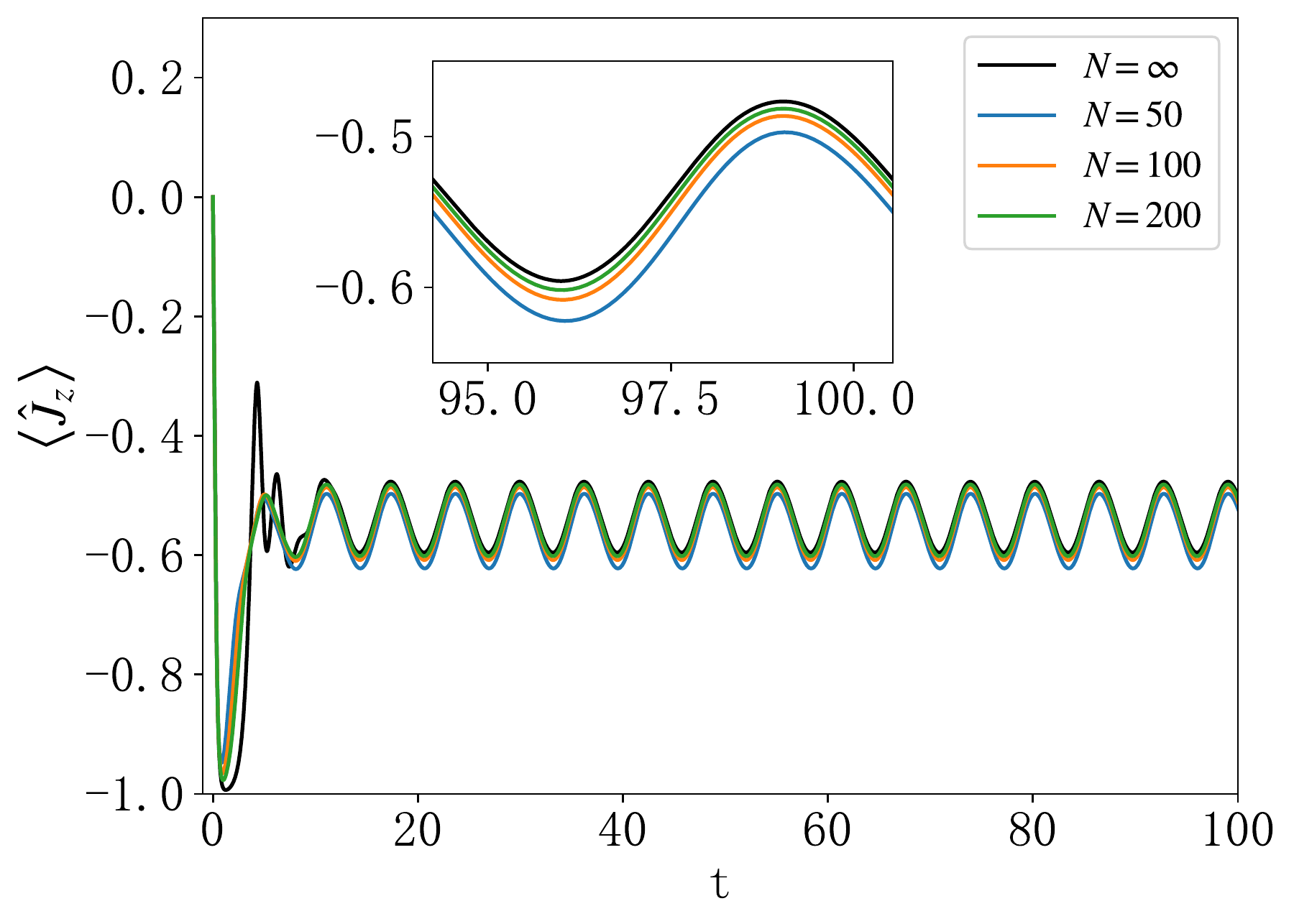}
		\end{minipage}\label{img4a}
	}%

	\subfigure[]{
		\begin{minipage}{1\linewidth}
			\centering
			\includegraphics[width=\linewidth]{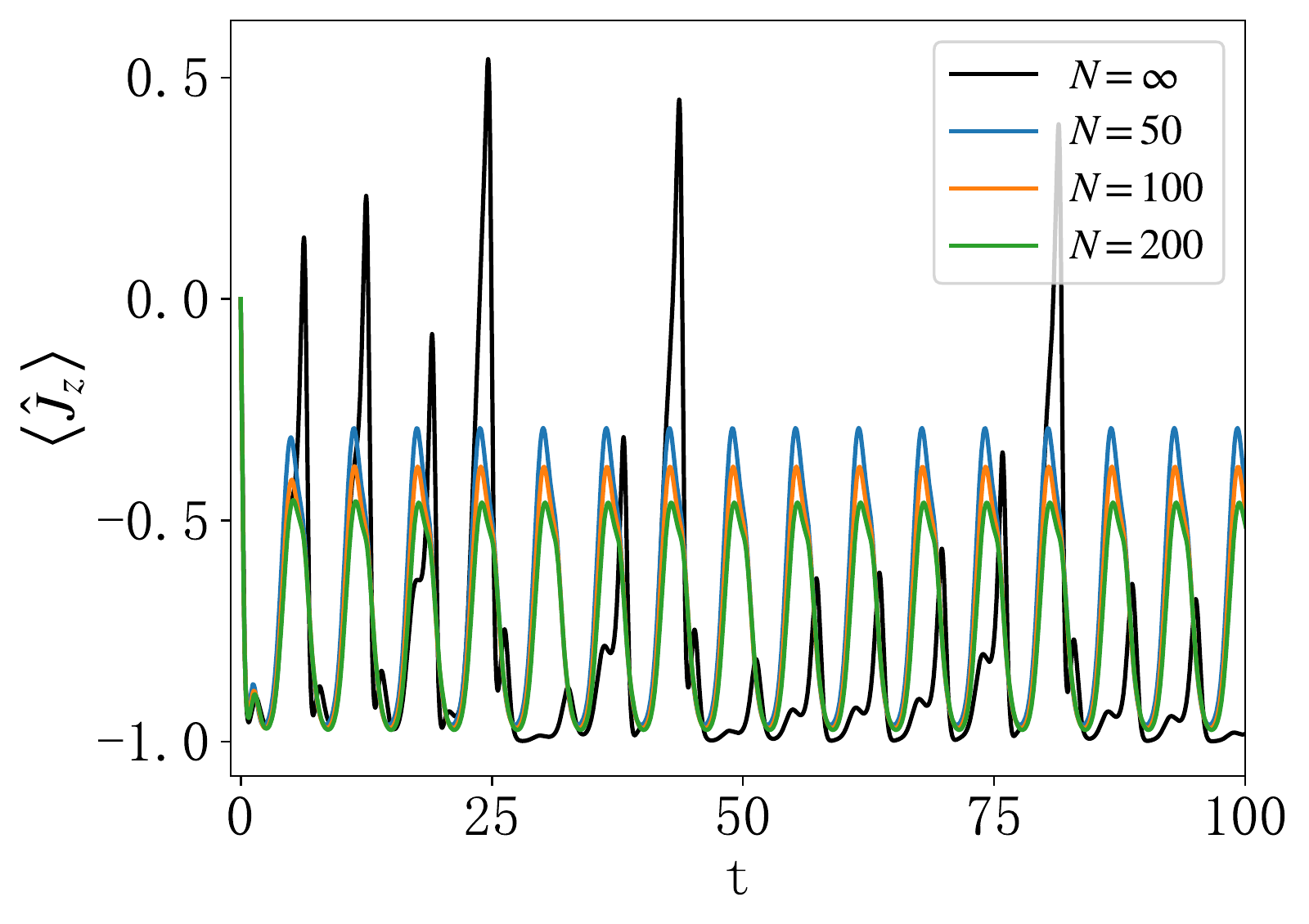}
		\end{minipage}\label{img4b}
	}%
	\centering
	\caption{The time evolution of $m_z$ for (a) $A=0.1$ and (b) $A=1$
	with different $N$. The black solid lines represent the semiclassical results.
	The initial condition is fixed to be $\theta_0=0.5\pi, \phi_0=0.1\pi$.}
	\label{img4(2)}
\end{figure}

In Fig.~\ref{img4a}, we compare the dynamics of magnetizations at
finite $N$s and in the semiclassical limit ($N=\infty$),
as $A=0.1$ is in the semiclassical periodic regime.
At finite $N$s, the magnetizations display periodic oscillations
in the asymptotic long time.
The curve of $m_z(t)$ at $N=50$ is already very close to the
semiclassical one. As $N$ increases, $m_z(t)$
goes even closer to the semiclassical result (see Fig.~\ref{img4a} the inset).
As $N\to\infty$, we expect that the numerical results should
coincide with the semiclassical results.
If $A$ is small, then the semiclassical approximation
is good for large enough $N$s.

Figure~\ref{img4b} shows the comparison as $A=1.0$
is in the semiclassical chaotic regime. The results
are significantly different from those at $A=0.1$. Up to $N=200$,
we find no similarity between the numerical result and the semiclassical one.
For $N$ ranging between $50$ and $200$, the initial transient
behavior of $m_z(t)$ always quickly evolves into the asymptotic behavior
- periodic oscillation. And the $m_z(t)$s for $N=50$ and $N=200$ have the same period,
with only their amplitude being different. But in the semiclassical result, $m_z(t)$ is aperiodic
at arbitrarily long time. These observations suggest that the
exact numerical solutions do not conserve to the semiclassical one
in the limit $N\to\infty$. If $A$ is large, then the semiclassical approximation is bad,
even for large $N$s. This seems to be strange for a fully-connected model.
We will further discuss this discrepancy in next section.

\begin{figure}
	\subfigure[]{
		\begin{minipage}{1\linewidth}
			\centering
			\includegraphics[width=\linewidth]{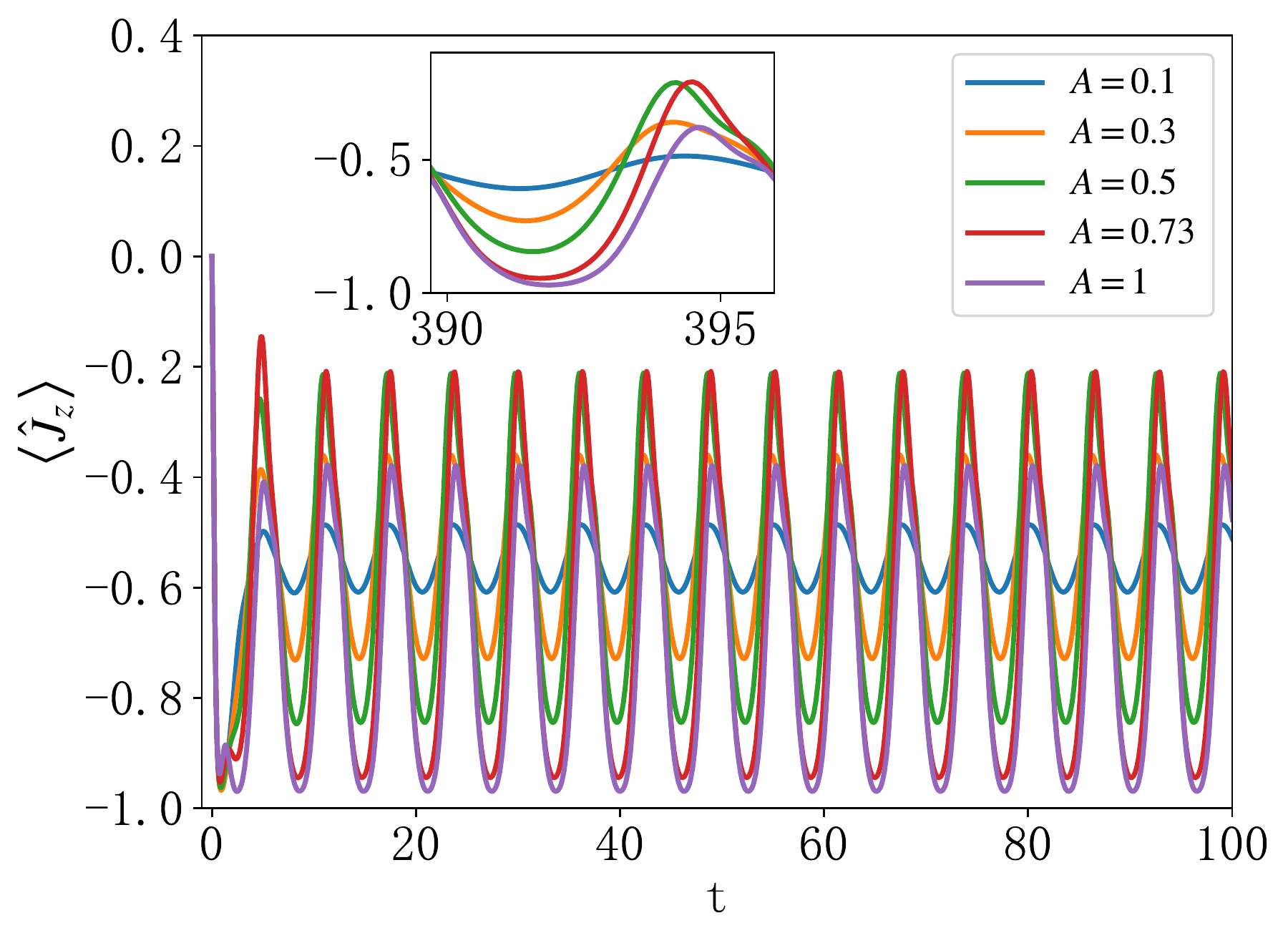}
		\end{minipage}\label{img5b}
	}%
	
	\subfigure[]{
		\begin{minipage}{1\linewidth}
			\centering
			\includegraphics[width=\linewidth]{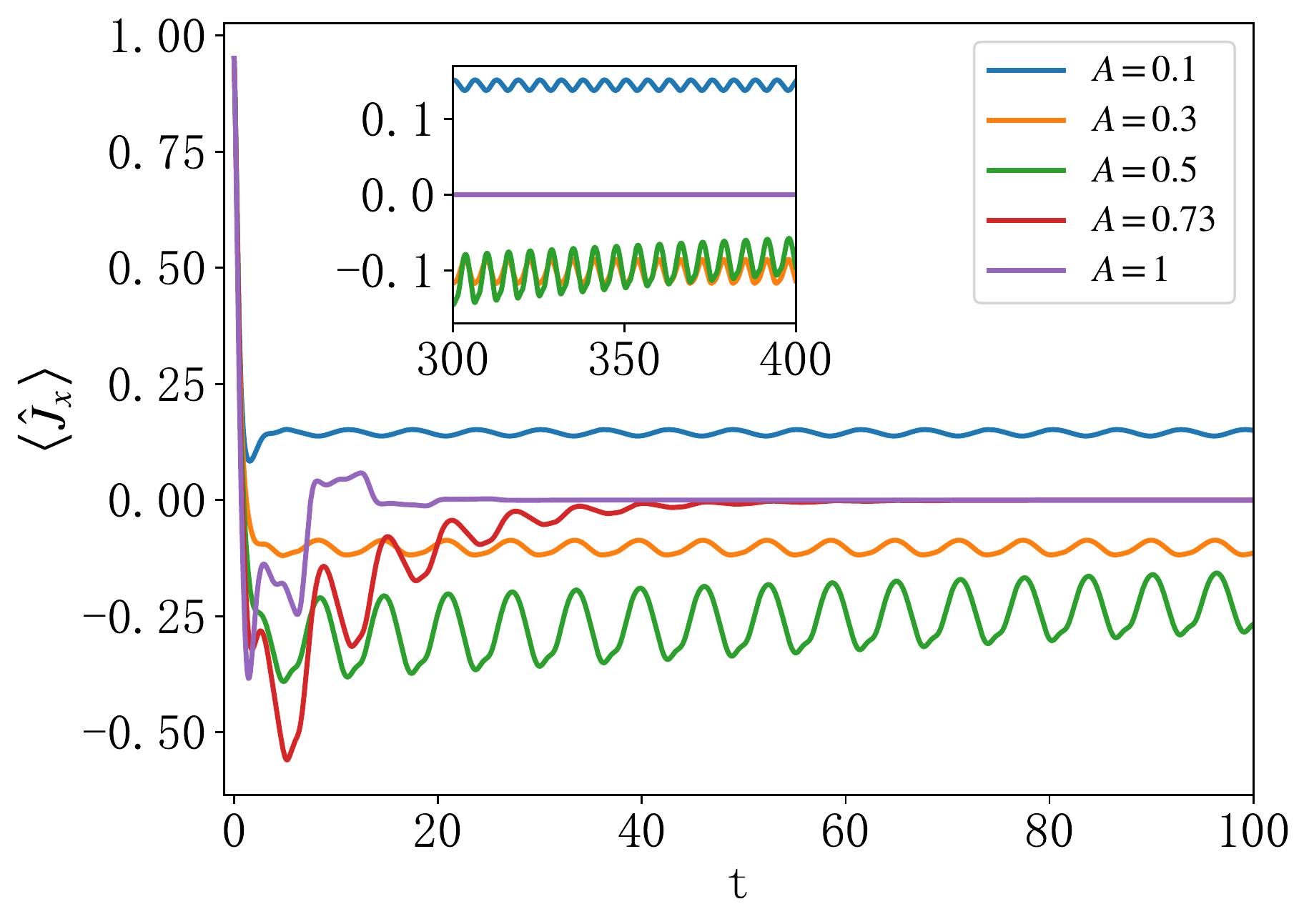}
		\end{minipage}\label{img5a}
	}%
	
	\centering
	\caption{The evolution of $m_x$ and $m_z$ with different $A$.
	The number of spins is chosen to be $N=100$. And the
	initial condition is $\theta_0=0.5\pi,\phi=0.1\pi$.}
	\label{img5}
\end{figure}
We also compare the dynamics of magnetizations for different $A$s, as
$N$ is fixed. Figure~\ref{img5b} and~\ref{img5a} display $m_z(t)$s and
$m_x(t)$s, respectively, for the values of $A$ ranging from $0.1$ to $1.0$.
The magnetization in the $z$-direction always displays a periodic oscillation.
And the oscillations for different $A$s are in phase (see Fig.~\ref{img5b} the
inset). On the other hand, the magnetization in the $x$-direction displays
two different dynamical modes, depending on the value of $A$.
As seen in Fig.~\ref{img5a}, as $A$ is small ($A=0.1, 0.3$),
$m_x$ displays an everlasting oscillation with the period $2\pi/\omega_0$.
But as $A$ is large ($A=0.73, 1.0$), $m_x$ rapidly decays to zero.
For an intermediate $A$ ($A=0.5$), $m_x$ maintains an oscillation for
a relatively long time, but the decay can still be clearly seen.
The decay of $m_x$ is a signal of the discrepancy between the exact
results and the semiclassical approximations, for the latter have non-decaying
magnetizations in all directions (see App.~\ref{sec:app}).

\section{Floquet Liouvillian spectrum}
\label{sec:spec}

In the closed systems, it is widely believed that the semiclassical
approximation becomes exact for the fully-connected models
if the number of spins goes to infinity.
In the open systems, a similar result was obtained~\cite{carollo2021exactness},
as the Liouvillian is time-independent. But in this paper, our numerical
results suggest that the semiclassical approximation is bad even in the
limit $N\to\infty$, if there is a big time-periodic term in the Liouvillian.
Since we can only obtain the exact results at finite $N$, a scaling
analysis is helpful for confirming our viewpoint. Next we give a
scaling analysis of the Floquet Liouvillian spectrum.

\begin{figure}
	
	\subfigure[]{
		\begin{minipage}{0.9\linewidth}
			\centering
			\includegraphics[width=\linewidth]{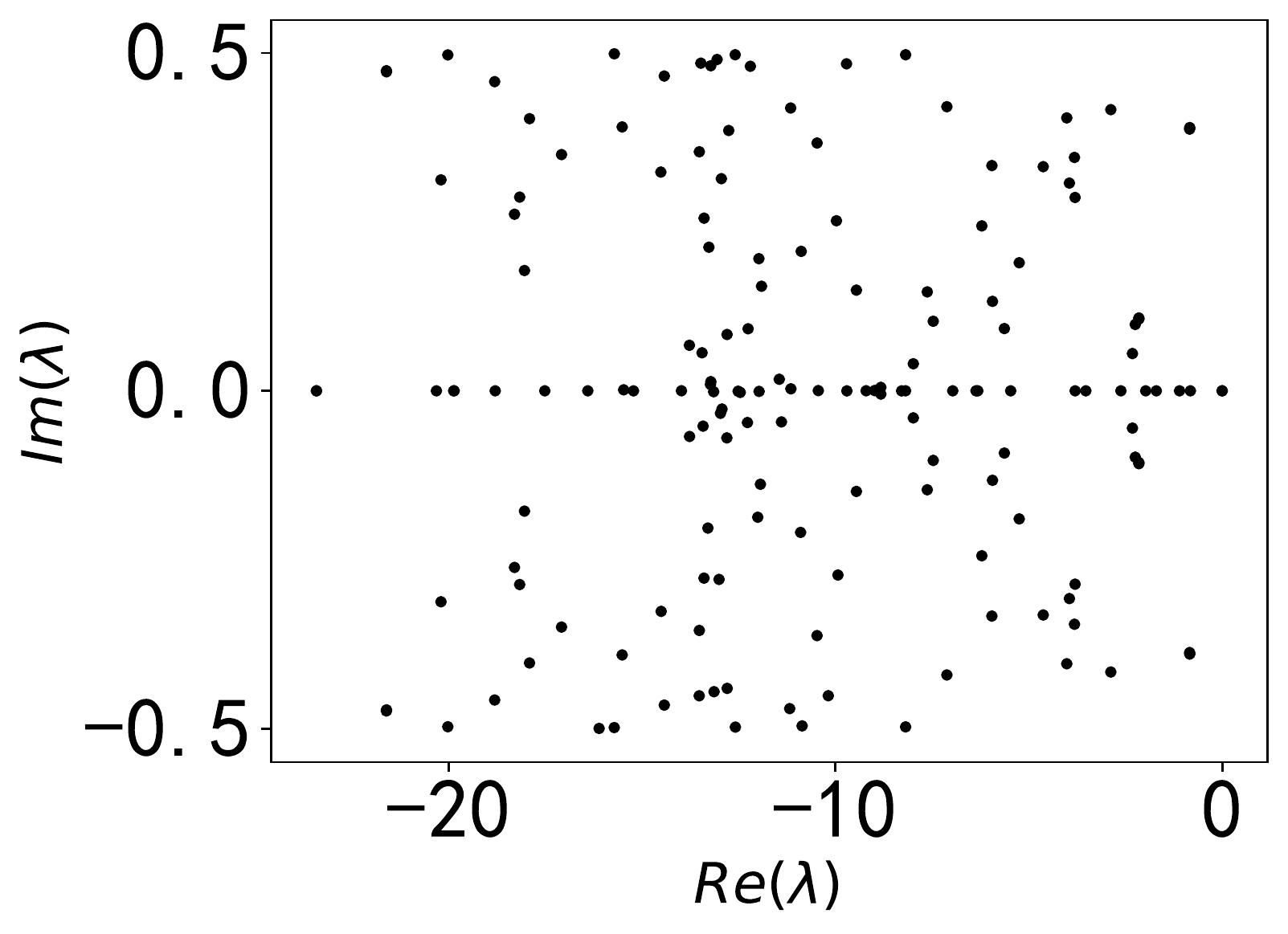}
		\end{minipage}\label{fig.6a}
	}%

	\subfigure[]{
		\begin{minipage}{0.9\linewidth}
			\centering
			\includegraphics[width=\linewidth]{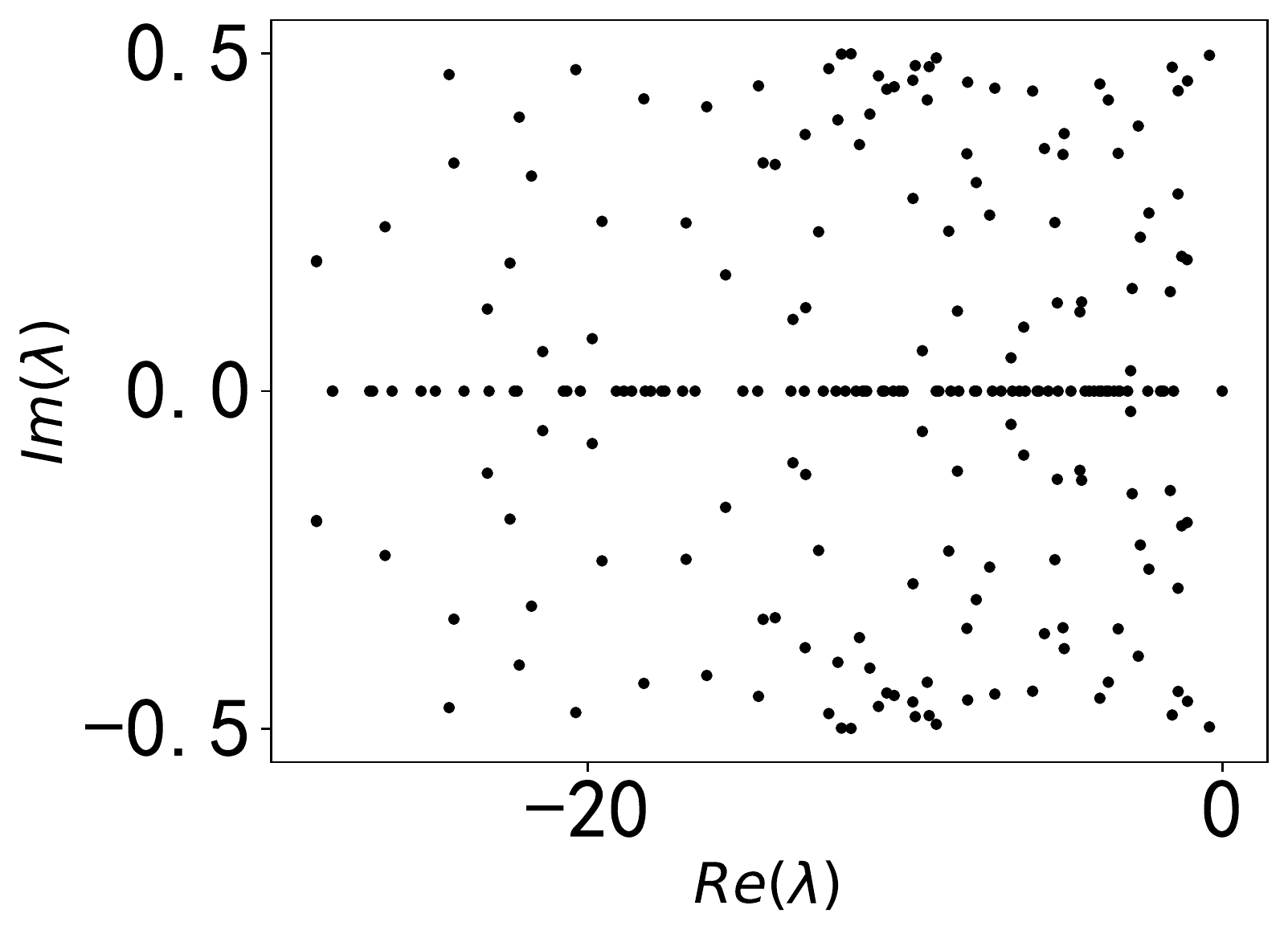}
		\end{minipage}\label{fig.6b}
	}%
	\centering
	\caption{The Floquet Liouvillian spectrum for (a) $A=0.1$ and (b) $A=1.0$.
	The number of spins is $N=30$.}
	\label{fig.6}
\end{figure}
The Lindblad equation can be expressed in a vectorized form as
$d\hat{\rho}/dt = \hat{\hat{\mathcal{L}}}\left( \hat{\rho}\right)$, where
$\hat{\hat{\mathcal{L}}}$ is the so-called Liouvillian
superoperator (or Liouvillian in short), which
is a non-Hermitian linear operator acting on the vector space of density matrices.
For the dissipative systems with time-independent
$\hat{\hat{\mathcal{L}}}$, it is well known that the eigenvalues
and eigenvectors of $\hat{\hat{\mathcal{L}}}$ determine the dynamics.
The eigenvalues of $\hat{\hat{\mathcal{L}}}$ (Liouvillian spectrum) are complex numbers.
More important, if the system size $N$ is finite, all the eigenvalues
must have negative real part, except one that is zero.

These notations can be generalized to the case of time-periodic $\hat{\hat{\mathcal{L}}}$.
For a Lindblad equation with $\hat{\hat{\mathcal{L}}}(t) = \hat{\hat{\mathcal{L}}}(t+T)$,
there exist a complete set of solutions written as $\hat{\rho}(t)= e^{\lambda t}  \hat{\varrho}(t)$,
where $\hat{\varrho}(t)=\hat{\varrho}(t+T)$ is the time-periodic part of density matrix,
according to the Floquet theorem. And $\hat{\varrho}(t)$ satisfies
\begin{equation}
\hat{\hat{\mathcal{L}}}\left( \hat{\varrho}(t) \right) - \frac{d}{dt} \hat{\varrho}(t)
= \lambda \hat{\varrho}(t).
\end{equation}
Then $\lambda$ can be seen as the eigenvalue of $\left(\hat{\hat{\mathcal{L}}}(t)- d/dt\right)$, which
is an operator acting on the generalized vector space of density matrices,
just as $\left(\hat{H}-id/dt\right)$ is an operator acting
on the generalized Hilbert space (Sambe space)
for a closed system with time-periodic Hamiltonian.
While the eigenstates of $\left(\hat{H}-id/dt\right)$ are called the
Floquet spectrum, the eigenstates of $\left(\hat{\hat{\mathcal{L}}}(t)- d/dt\right)$
are called the Floquet Liouvillian spectrum.

Compared to the Floquet spectrum or Liouvillian spectrum, much less is known about the
Floquet Liouvillian spectrum. For the dissipative systems, we guess that
the Floquet Liouvillian spectrum has the same properties as the Liouvillian spectrum,
i.e., all the complex eigenvalues have negative real parts except for a unique one
that is zero. The numerics support our guess. Figure~\ref{fig.6}
displays the Floquet Liouvillian spectrum at $N=30$. For both
$A=0.1$ and $A=1.0$, we see that the rightmost eigenvalue
in the complex plane is zero, which is non-degenerate.
And the others are to the left of zero.

The Floquet Liouvillian spectrum completely determines whether
the dynamics is periodic, subharmonic or chaotic. To see it, we arrange all
the eigenvalues as
\begin{equation}
\lambda_0=0\geq \text{Re}\lambda_1 \geq \text{Re}\lambda_2 \geq \cdots,
\end{equation}
with the corresponding eigenvectors being $\hat{\varrho}_0(t)$, $\hat{\varrho}_1(t)$,
$\hat{\varrho}_2(t),\cdots$. For an arbitrary initial state, the
solution of Lindblad equation can be formally expressed as
\begin{equation}
\hat{\rho}(t) = \sum_j K_j e^{\lambda_j t}  \hat{\varrho}_j(t),
\end{equation}
where the coefficients $K_j$ depend on the initial state.
In the asymptotic long time, the terms with $\text{Re}\lambda_j<0$ all
decay to zero, and $1/\left| \text{Re}\lambda_j \right|$ is just the decay time of
the $j$-th mode. At finite $N$, all the $\lambda_j$ with $j>0$ have
negative real parts, therefore, the density matrix in the asymptotic long time
becomes $\hat{\rho}(t) = K_0  \hat{\varrho}_0(t)$, which is exactly periodic with
the period $T$. We then expect that the asymptotic behavior is always periodic at finite $N$.
The situation is more complicated in the thermodynamic limit.
As found in the previous studies of time-independent Liouvillians,
there exist possibilities that the real parts of some $\lambda_j$ decrease
with increasing $N$ and vanish in the limit $N\to\infty$. As a consequence, the asymptotic
density matrix becomes $\hat{\rho}(t) = \sum_{j=0}^L K_j 
e^{i \text{Im}\left(\lambda_j\right) t}  \hat{\varrho}_j(t)$, where $L$ is the number of
eigenvalues with vanishing real parts. Additionally, if these $\lambda_j$
have also nonvanishing imaginary parts, then $\hat{\rho}(t)$ possibly display
subharmonic or chaotic behaviors, depending on the values of $\text{Im}\left(\lambda_j\right)$.
Conversely, if the real parts of $\lambda_j$ with $j>0$ are all
finite in the limit $N\to\infty$, then subharmonic oscillation or chaotic
behavior are both impossible.

\begin{figure}
	\centering
	\includegraphics[width=1\linewidth]{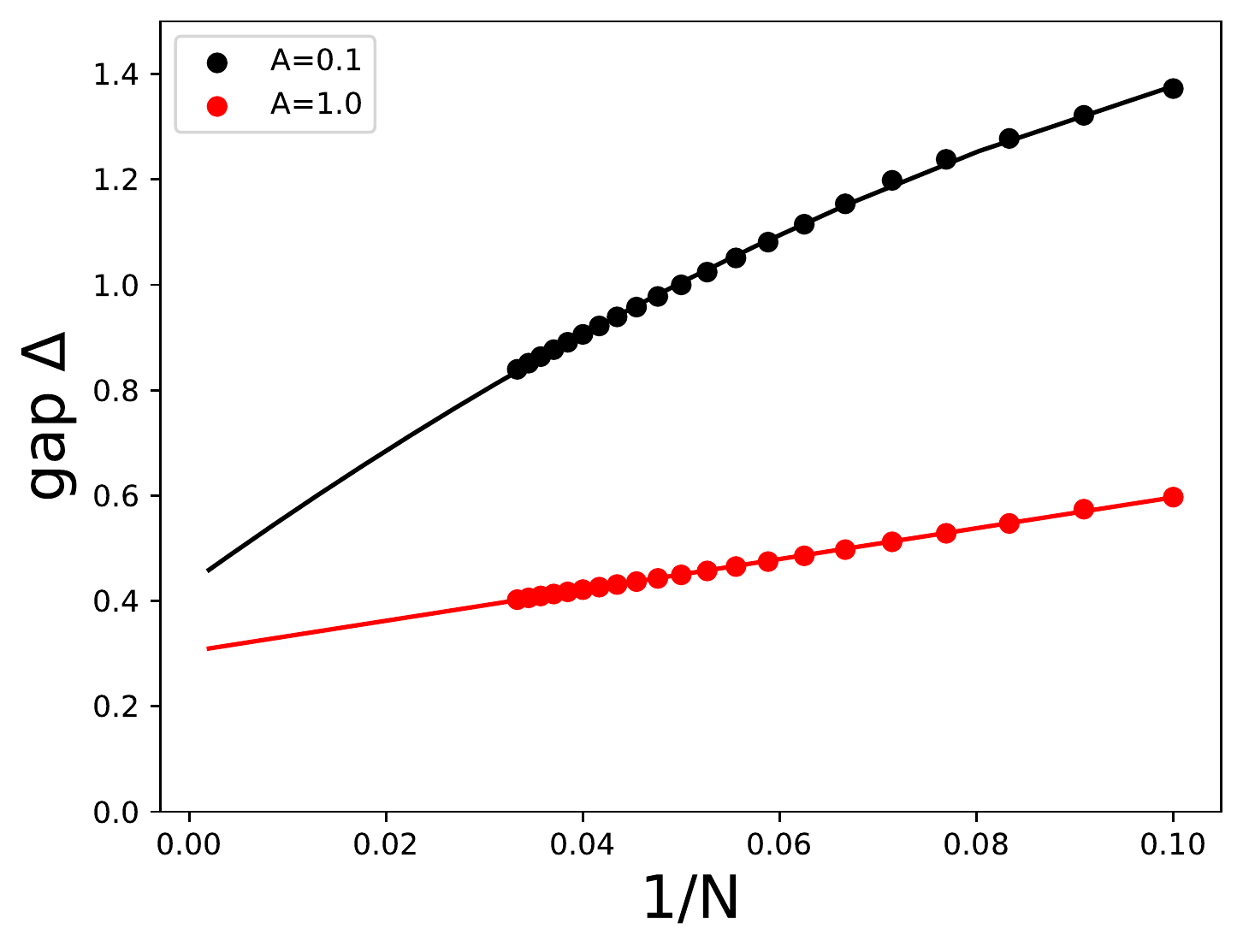}
	\caption[]{The Floquet Liouvillian gap as a function of $1/N$. The dots
	represent the numerical results, while the lines represent the fitted curves.}
	\label{fig.7}
\end{figure}
According to the above argument, we perform a scaling analysis of
the Floquet Liouvillian gap, which is defined as $\Delta = \lambda_0 - \text{Re}\lambda_1$.
Figure~\ref{fig.7} plots $\Delta$ as a function of $1/N$.
The dots are the numerical results, while the lines are the fitted curves.
The gap at $A=1.0$ is significantly smaller than the gap at $A=0.1$.
But in both cases, we clearly see that the gap does not go
to zero as $N\to\infty$. This indicates that no $\lambda_j$ with $j>0$
has vanishing real parts, therefore, the asymptotic behaviors
are periodic for both $A=0.1$ and $A=1.0$, even in the limit $N\to \infty$. Such a result is
consistent with our previous simulation of the real-time dynamics of magnetizations.

\section{CONCLUSIONS}
\label{sec:con}

In summary, we study the fully-connected Ising model with
a time-periodic external field and subject to a dissipation,
by using both the semiclassical approach and the exact
numerical simulation. If the field amplitude is small, both the semiclassical approach
and the numerical simulation predict a perfect periodic oscillation
of magnetizations. And the numerical results in the thermodynamic limit
are consistent with the semiclassical one.

As the field amplitude increases, the semiclassical approach predicts
the period-doublings or subharmonic oscillations, and a series of period-doublings
finally lead to the chaotic dynamics of magnetizations, which is confirmed
by the calculations of Lyapunov exponents. On the contrary, the
numerical simulation show that the magnetizations are always
oscillating periodically, whatever the field amplitude is. No subharmonic
or chaotic dynamics are observed, even we choose the 
number of spins to be as large as a few hundreds in the simulation.
We then analyze the Floquet Liouvillian gap, which conserves to a finite
value in the thermodynamic limit, for either small or large field amplitude.
We argue that a finite gap is another evidence of the periodic oscillations of observables.

For the fully-connected models, the
semiclassical approximation is generally believed to be good
for sufficiently large number of spins. But we find that this is
not the case if the time-periodic field and the dissipation are both present.
For a large field amplitude, the predictions from the semiclassical approach
and the numerical method are qualitatively different.
Our finding suggests that one should be more careful when
using the semiclassical approach in the case of time-periodic Liouvillians.

\section*{Acknowledgement}
This paper is supported by NSFC under Grant Nos.~11774315,~11835011, and~12174346,
and by the Junior Associates program of the Abdus Salam International Center for Theoretical Physics.

\appendix

\section{Trajectory on the Bloch sphere}
\label{sec:app}

\begin{figure}
	\centering
	\includegraphics[width=1\linewidth]{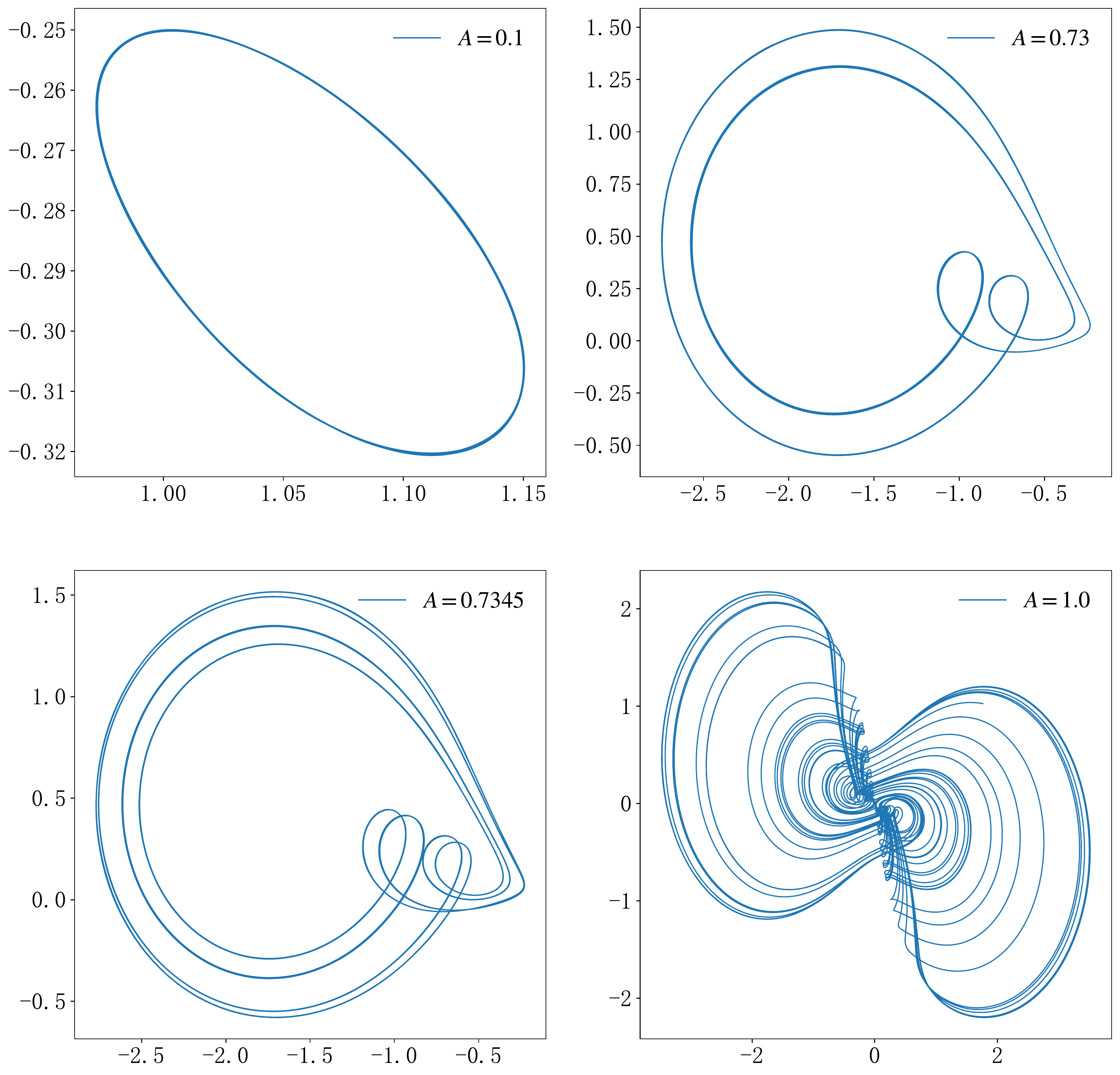}
	\caption[]{The trajectories of ${\bf{m}}(t)$ during $t\in \left[16T_0,80T_0\right]$
		with $T_0=2\pi/\omega_0$ being the time period, for $A=0.1,0.73,0.7345$ and $1.0$.
		The initial condition is chosen to be $(\theta ,\varphi )=(0.5\pi,0.1\pi)$.}
	\label{img9}
\end{figure}

Because $\left| {\bf{m}}(t)\right| $ is a constant of motion in the semiclassical approach,
the vector ${\bf{m}}=(m_x, m_y, m_z)$ is moving on a Bloch sphere.
Without loss of generality, we set $\left| {\bf{m}}\right|=1$.
For $\left| {\bf{m}}\right|\neq 1$, we can always
rescale $\left| {\bf{m}}\right|$ to unity by changing the units.
To better display the trajectory of ${\bf{m}}(t)$ on a sphere, we perform the stereographic
projection and map the unit sphere into the $x-y$ plane.
The map is defined by $x=2m_x/(1-m_z) $ and $y=2m_y/(1-m_z)$.

Figure~\ref{img9} displays the trajectories in the $x$-$y$ plane for
$A=0.1,0.73,0.7345$ and $1.0$. We choose the time interval to be
$\left[16T_0,80T_0\right]$ with $T_0=2\pi/\omega_0$ being the period.
The periodic, subharmonic and chaotic behaviors are clearly distinguishable.
As $A=0.1$, the trajectory is periodic with an oval shape.
As $A=0.73$, the trajectory has two different loops, which is a signature
of period doubling, as it should be.
As $A=0.7345$, the trajectory has four different loops, indicating that
the period is four times of the original one.
As $A=1.0$, we find the trajectory is aperiodic, showing features of chaos.
From Fig.~\ref{img9}, we can also see that the three components are all
nonzero, otherwise, the trajectory in the plane should be a circle or straight line.
\bibliography{ref.bib}

\end{document}